# Multi-channel quantum noise suppression and phase-sensitive modulation in a hybrid optical resonant cavity system

Ke Di[1], Shuai Tan [1], Liyong Wang[2,*], Anyu Cheng[1], Xi Wang[1], Yuming Sun[1], Junqi Guo[1], Yu Liu[1], and Jiajia Du[1,*]

[1]Chongqing University of Post and Telecommunications, Chongqing, 400065, China
[2]Department of Applied Physics, Wuhan University of Science and Technology, Wuhan 430081, China

E-mail: *wangliyong@wust.edu.cn*
    *dujj@cqupt.edu.cn*

**Abstract**

Quantum noise suppression and phase-sensitive modulation for continuous variable of vacuum and squeezed fields in a hybrid resonant cavity system are investigated theoretically. Multiple dark windows similar to electromagnetic induction transparency (EIT) are observed in quantum noise fluctuation curve. The effects of pumping light on both suppression of quantum noise and control the widths of dark windows are carefully analyzed, and the saturation point of pumping light for nonlinear crystal conversion is obtained. We find that the noise suppression effect is very sensitive to the pumping light power. The degree of noise suppression can be up to 13.9dB when the pumping light power is $0.65\beta_{th}$. Moreover, a phase-sensitive modulation scheme is demonstrated, which well fills the gap that multi-channel quantum noise suppression is difficult to realize at the quadrature amplitude of squeezed field. Our result is meaningful for various applications, especially in precise measurement physics, quantum information processing and quantum communications of system-on-a-chip.

Keywords: Quantum Information Processing, Quantum Communication, Weak Signal Measurement

## 1. Introduction

Quantum interference is a hot and important research topic in fields of quantum physics and optical physics. Squeezed light is an essential component of a continuous variable (CV) in quantum information processing [1, 2]. The quantum interference phenomenon based on squeezed light has been studied extensively over the past decades [3-5]. At present, the squeezed degree of light field based on optical parametric amplification (OPA) technique is significantly high, and it is up to 15dB [6]. Squeezed light can be used in various applications like measuring weak light signal [7], detecting gravitational waves [8, 9], testing fundamental physics [10], realizing quantum information processing and optical communications [11, 12], etc. In year of 2020, Frascella G et al. reported that a squeezing-assisted interferometer with phase sensitivity overcame 3dB shot noise limit [13]. This result improves the interferometer performance significantly and decreases the detection loss rate greatly. In the same year,

Zuo X J et al. showed that the phase sensitivity of quantum interferometer can be improved to 4.86 ± 0.24 dB via using the squeezed state of light field and amplifying the phase sensing intensity [14]. From the above discussion, it can be known that the phase-sensitive manipulation of squeezed state light field is of great importance and has remarkable potential to improve the quantum noise suppression degree [15].

Electromagnetic induced transparency (EIT) is a quantum interference phenomenon between absorbing medium and light fields [16-19]. In 2020, broadband coherent optical storage based on EIT protocol was proposed by Wei Y C et al., and their storage efficiency is about 80% with a pulse duration of 30ns [20, 21]. In 2021, a scheme for storage and retrieval of optical Peregrine solitons via EIT was proposed by Shou C et al. Optical Peregrine solitons with very small propagation loss, ultraslow motional velocity, and extremely low generation power were created [22]. Despite the above approaches which demonstrated the excellent characteristics of EIT in terms of quantum storage, a simultaneous realization of multifaceted storage and fast storage time is still not well solved. In 2021, Pan J W's group successfully developed a prototype of Ninth Chapter II quantum computer [23, 24]. This breakthrough increased the number of multi-photon quantum interference lines from 100 to 144 dimensions, and the number of photons manipulated has improved from 76 to 113 [25, 26]. However, their system was realized in the framework of dissociated variables (DV). Comparing to DV, it is well known that CV has superiority potential both for higher detecting efficiency and transmitting bit efficiency [27, 28]. Here we propose a multi-channel noise suppression scheme which will help to improve the quantum storage efficiency and phase-sensitive manipulations of CV, thus it will has important applications in quantum information processing [29], quantum entanglement [30-34], quantum imaging [35], optical deceleration [36], etc.

In this paper, we propose a scheme for multi-channel quantum noise suppression at quadrature amplitude of vacuum field and squeezed field in a hybrid resonant cavity system with phase-sensitive modulation. A series of EIT-like dark windows in quantum noise fluctuation curve are produced since the strong-coupling between cavity fields and the periodically-poled potassium titanyl phosphate (PPKTP) crystal. As far as we know, this is the first time that EIT-like phenomena are generated simultaneously at the resonance frequency ($\Delta_s = 0$ MHz) and sideband detuning frequencies ($\Delta_s = \pm 3.1$ MHz) in noise curve. The system can be easily manipulated by shifting the phase of the pumping field to achieve phase-sensitive quantum modulation. Furthermore, a detailed comparison of the OPA resonant cavity and the hybrid resonant cavity is given. The saturation point of nonlinear crystal conversion is analyzed, and the maximum degree of the quadrature amplitude noise suppression with phase-sensitive modulation is obtained. It shows that the degree of noise suppression is very sensitive to the pumping light power. When the pumping power is increased to extreme value $0.65\beta_{th}$, the degree of noise suppression is up to 13.9dB.

## 2. The model

As shown in Fig. 1(a), a hybrid optical resonant cavity consists of two flat-concave mirrors $M_{front}$, $M_{bac}$, and a plane mirror $M_{mid}$. The plane mirror $M_{mid}$ is at the center of the resonant cavity, and it divides the coaxial optical cavity into two optical cavities $C_1$ and $C_2$ with nonlinear coupling effect on each cavity (the front cavity is represented by $C_1$, the back cavity is represented by $C_2$ and two cavities have the same cavity resonant frequency). The coupling coefficient depends on the reflectivity rate of the middle cavity mirror. A Periodically PPKTP crystal is placed at the center of cavity $C_2$. Optical cavity $C_2$ in Fig. 1(a) consists of a flat-concave mirror $M_{bac}$, a plane mirror $M_{mid}$ and a nonlinear crystal, which forms an OPA system. The OPA in our study works below threshold. The annihilation of a pumping photon at high frequency in cavity $C_2$ produces a low-frequency signal photon and an idle photon. As a result, the number of signal photons increases, which leads to the light amplification effect.

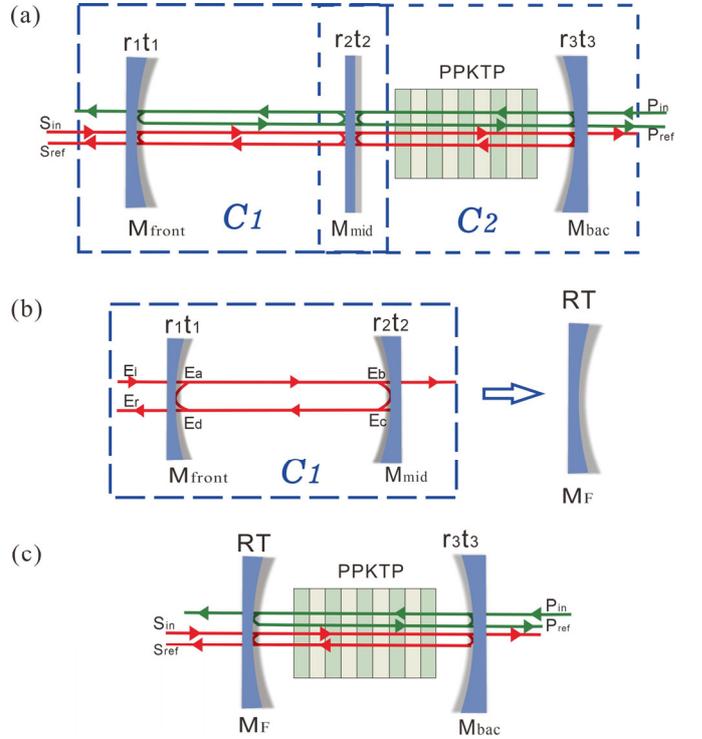

FIG. 1. Schematic diagram of the hybrid optical resonant cavity system. (a) Optical cavity $C_1$ connects optical cavity $C_2$ with a plane mirror $M_{mid}$. $r_1$, $r_2$ and $r_3$ denote the reflection coefficients and $t_1$, $t_2$ and $t_3$ denote the transmission coefficients of the signal light in cavity mirrors $M_{front}$, $M_{mid}$ and $M_{bac}$, respectively. $V_{in}$, $V_{ref}$ denote the input and reflected fields of light respectively where $V = S, P$. $S$ is the signal light. $P$ is the pumping light. (b) Equivalent model of cavity $C_1$. $E_a$, $E_b$, $E_c$, $E_d$ denote light field amplitude in different positions of cavity $C_1$, $E_i$, $E_r$ denote incident light and reflected light, respectively. $R$ and $T$ are the total reflectance and general transmittance of cavity $C_1$. (c) Simplified equivalent model of the hybrid optical parametric amplification cavity.

The Hamiltonian of the system can be written as [37-40]:

$$H = \hbar\omega_p \hat{a}_p^+ \hat{a}_P + \hbar\omega_s \hat{a}_s^+ \hat{a}_s + \hbar\omega_i \hat{a}_i^+ \hat{a}_i$$
$$+ i\hbar g\left(\hat{a}_s^+ \hat{a}_i^+ \hat{a}_p - \hat{a}_s \hat{a}_i \hat{a}_p^+\right)$$
$$+ i\hbar \left(E_p \hat{a}_p^+ e^{-i\omega_0^p t} + E_s \hat{a}_s^+ e^{-i\omega_0^s t} + H.c.\right) \quad (1)$$

where the first three terms on the right side of Eq. (1) represent the free Hamiltonian of the pumping field, signal field and idle field, respectively. $\hat{a}_j$ ($\hat{a}_j^+$) ($[\hat{a}_j, \hat{a}_j^+] = 1$, $j = p, s, i$) is the annihilation (creation) operator of the pumping, signal and idle modes, respectively. Correspondingly, $\omega_p$, $\omega_s$ and $\omega_i$ are frequencies of the pumping, signal and idle modes. The fourth term in Eq. (1) denotes the interaction of three laser fields, where the pumping light performs parametric down-conversion in the nonlinear crystal to generate signal light and idle light. $g$ is the nonlinear susceptibility. The last term in Eq. (1) describes the input driving by the pumping field and the signal field with frequency $\omega_0^p$, $\omega_0^s$, respectively. The input driving term does not include the idle field since the idle field is converted from the pumping field in a nonlinear crystal. $E_n = \sqrt{2P_n\gamma_n/\hbar\omega_0^n}$ ($n = p, s$) denotes the coupling strength between the optical cavity and each light field. $P_n$ is the power of light field with frequency $\omega_0^n$, and $\gamma_n = \gamma_f^n + \gamma_0^n + \gamma_{out}^n$ is the total decay rate of the pumping and signal modes in the cavity. Here $\gamma_f^n$ ($\gamma_{out}^n$) is the decay rate of light field $\hat{a}_n$ going through the front and middle (back) cavity mirrors, $\gamma_0^n$ is intracavity decay of light field in hybrid cavity system.

The hybrid cavity system can produce multi-channel quantum interference effect compared with a single-cavity (without the plane mirror $M_{mid}$). Based on the amplitude of the input and reflected fields at different positions in cavity $C_1$ (Fig. 1(b)), the overall reflection coefficient $R$ of cavity $C_1$ can be derived as [41]:

$$R = \frac{E_r}{E_i} = \frac{-r_2 + \gamma_c r_1 \exp\left(\frac{2iwL}{c}\right)}{1 - \gamma_c r_1 r_2 \exp\left(\frac{2iwL}{c}\right)} \quad (2)$$

Therefore, the hybrid resonant cavity system can be simplified to an equivalent optical cavity model. $L$ is the length of this equivalent cavity, $w$ is the cavity frequency and $\gamma_c$ is other decay effects in cavity $C_1$. (Fig. 1(c)).

Consider the dissipation and input noise of each mode, the quantum Langevin equations describing the system are given by [42, 43]:

$$\dot{\hat{a}}_s = -i\Delta_s \hat{a}_s - \gamma_s \hat{a}_s + g\hat{a}_i^+ \hat{a}_p + E_s + \hat{a}_s^{In} \quad (3a)$$

$$\dot{\hat{a}}_p = -i\Delta_p \hat{a}_p - \gamma_p \hat{a}_p - g\hat{a}_s \hat{a}_i + E_p + \hat{a}_p^{In} \quad (3b)$$

$$\dot{\hat{a}}_i = -i\Delta_i \hat{a}_i - \gamma_i \hat{a}_i + g\hat{a}_s^+ \hat{a}_p + \hat{a}_i^{In} \quad (3c)$$

where $\Delta_n = \omega_n - \omega_0^n$, since the frequency of the idle field is equal to the signal field, so $\Delta_s = \Delta_i$. The input field fluctuation is defined as $\hat{a}_j^{In} = \sqrt{2\gamma_{in}^j}\hat{a}_j^{in} + \sqrt{2\gamma_{out}^j}\hat{a}_v^j$. The correlation function of the input optical field $\hat{a}_j^{in}$ and the vacuum field $\hat{a}_v^j$ are described by $\langle Q(t)Q^+(t')\rangle = [N+1]\delta(t-t')$, $\langle Q^+(t)Q(t')\rangle = N\delta(t-t')$, $\langle Q(t)Q(t')\rangle = M\exp[-i\Delta_j(t+t')]\delta(t-t')$ and $\langle Q^+(t)Q^+(t')\rangle = M^*\exp[i\Delta_j(t+t')]\delta(t-t')$, where $Q(t) = \hat{a}_j^{in}$ or $\hat{a}_v^j$. $N = \sinh^2 s$ and $M = \exp(i\theta)\sinh s \cosh s$. $s$ is the squeezing parameter of squeezed vacuum field, when $Q(t) = \hat{a}_v^j$, the squeeze index $s = 0$ [44]. The pumping field is $\hat{a}_p = \beta \exp(i\theta)$. $\beta$ and $\theta$ denote the amplitude and phase of the pumping light. The frequency of the idle field is equal to signal field when the system is fully resonant. Therefore, the quantum Langevin equation of motion based on the signal field can be simplified as [45]:

$$\dot{\hat{a}}_s = (-i\Delta_s - \gamma_s)\hat{a}_s + g\beta \exp(i\theta) \hat{a}_s^+ + \sqrt{2\gamma_{in}^s}\hat{a}_s^{in} + \sqrt{2\gamma_{out}^s}\hat{a}_v^s \quad (4)$$

where $\gamma_{in}^s = \gamma_f^s + \gamma_c = \sqrt{1-R^2}$ denotes the decay rate of signal field $\hat{a}_s$ going through the front cavity mirror $M_{front}$, the middle cavity mirror $M_{mid}$ (cavity $C_1$) and other decay effects in cavity $C_1$ [41].

The operator $\hat{a}_s$ in Eq. (3) contains a steady-state mean term at a certain point and a small quantum fluctuation term, i.e., $\hat{O} = \bar{O} + \delta\hat{O}$ ($\hat{O} = \hat{a}_s, \hat{a}_s^+$). The quadrature amplitude and phase of the signal field are $\delta\hat{X} = (\delta\hat{a}_s + \delta\hat{a}_s^+)/\sqrt{2}$ and $\delta\hat{Y} = i(\delta\hat{a}_s^+ - \delta\hat{a}_s)/\sqrt{2}$, respectively. Combining $\hat{X}$ and $\hat{Y}$ with the input-output theory $\delta\hat{a}_{re} = -\hat{a}_s^{in} + \sqrt{2\gamma_{in}^s}\delta\hat{a}_s$, after a Fourier transform we get the reflected field noise covariance of quadrature amplitude:

$$\delta^2 X(\omega) = \frac{P_X \delta^2 X_{in}(\omega) + Q_X \delta^2 Y_{in}(\omega) + M_X \delta^2 X_v(\omega) + N_X \delta^2 Y_v(\omega)}{[\gamma_s^2 + (\Delta_s \tau)^2 - (\tau\omega)^2 - (g\beta)^2]^2 + 4(\tau\omega\gamma_s)^2} \quad (5)$$

Similarly, the noise covariance of quadrature phase is obtained as:

$$\delta^2 Y(\omega) = \frac{P_Y \delta^2 Y_{in}(\omega) + Q_Y \delta^2 X_{in}(\omega) + M_Y \delta^2 Y_v(\omega) + N_Y \delta^2 X_v(\omega)}{[\gamma_s^2 + (\Delta_s \tau)^2 - (\tau\omega)^2 - (g\beta)^2]^2 + 4(\tau\omega\gamma_s)^2} \quad (6)$$

where the parameters $P_k$, $Q_k$, $M_k$, $N_k$ are given by:

$$P_k = [\gamma_s(2\gamma_{in} - \gamma_s) + (\tau\omega)^2 - (\Delta_s\tau)^2 + (g\beta)^2 \pm 2g\beta\gamma_{in}\cos\theta]^2 + 4(\tau\omega)^2(\gamma_s - \gamma_{in})^2 \quad (7a)$$

$$Q_k = 4\gamma_{in}^2(\Delta_s\tau \pm g\beta\sin\theta)^2 \quad (7b)$$

$$M_k = 4\gamma_{in}\gamma_{out}[(\gamma_s \pm g\beta\cos\theta)^2 + (\tau\omega)^2] \quad (7c)$$

$$N_k = 4\gamma_{in}\gamma_{out}(\Delta_s\tau \pm g\beta\sin\theta)^2 \quad (7d)$$

Eqs. (7a-7d) get plus signs when $k = X$, and they get minus signs when $k = Y$. $\delta^2 Y_{in}(\omega) = \exp(-2s)$ and $\delta^2 X_{in}(\omega) = \exp(2s)$ denote noise fluctuation. The mean value of vacuum noise covariance is $\delta^2 X_v(\omega) = \delta^2 Y_v(\omega) = 1$.

## 3. Analysis and discussion

### 3.1 Vacuum fields

Firstly, we focus on the multi-channel quantum noise suppression of vacuum field. The squeezing index is 0 in this case. Single-cavity model can be achieved by removing the cavity mirror $M_{mid}$ of the system. The transmission coefficient of the front and back cavity mirrors are $t_1 =$ 0.0016 and $t_3 = 0.00005$ [46, 47]. Fig. 2 shows the quantum noise curves with the pumping power $0.2\beta_{th}$ and $0.5\beta_{th}$. The threshold limit $\beta_{th} = (\gamma_p\sqrt{\gamma_s\gamma_i})/g$ [40]. In Fig. 2(a), as the pumping filed interacting with the degenerate OPA cavity, the quantum noise is amplified at resonance frequency $\Delta_s = 0$ MHz and two weak squeezed states appear at sideband detuning frequencies symmetrically. The coupling strength of nonlinear PPKTP crystal and optical cavity is enhanced as the pumping power increases, which leads to deeper squeezed depth at two sidebands. The quadrature amplitude shows an amplified state evidently due to the interaction between the pumping light and the nonlinear crystal under the Heisenberg uncertainty principle. Correspondingly, the quadrature phase shows a squeezed state in quantum noise curve in Fig. 2(b). The squeezed depth of quadrature phase is enhanced as the pumping power increases, and the noise suppression degree reaches 8.6dB when the pumping power is $0.5\beta_{th}$.

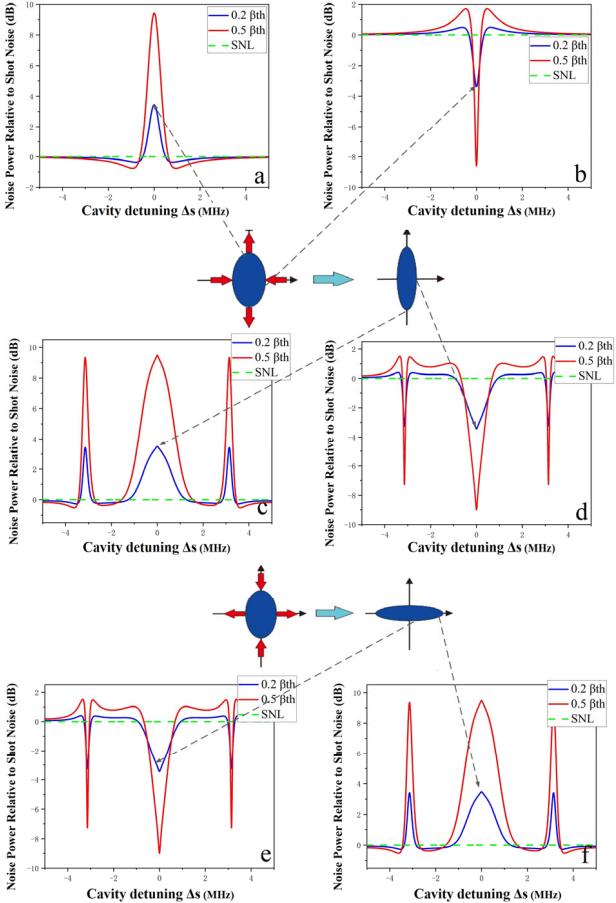

FIG. 2. The quantum correlation noise fluctuation curve of vacuum field in a hybrid resonant cavity system. Green curve: shot-noise limit (SNL); Blue and red curves: the pumping field powers are $0.2\ \beta_{th}$ and $0.5\ \beta_{th}$, respectively. (a) and (b) denote noise fluctuation curves for quadrature amplitude and phase in a single-cavity. (c) and (d) denote noise curves for quadrature amplitude and phase in a hybrid resonant cavity with $\theta = 0$. (e) and (f) denote noise curves for quadrature amplitude and phase in a hybrid resonant cavity with $\theta = \pi$.

Next, consider the vacuum field in a hybrid optical cavity system with three cavity mirrors $M_{front}$, $M_{mid}$ and $M_{bac}$. The coupling strength between cavities $C_1$ and $C_2$ depends on the transmissivity $t_2$. According to the accessible parameters in experiments by Di K et al. [15], we set transmission coefficients of cavity mirrors as $t_1 = 0.016$, $t_2 = 0.26$ and $t_3 = 0.002$. As shown in Fig. 2(c) and Fig. 2(d), multiple interference channels in quantum noise curve are obtained due to the coupling of cavities $C_1$ and $C_2$. These quantum interference channels are very sensitive to the pumping light parameters. The squeezed degree enhances significantly as the pumping light power increases. In particularly, the quadrature phase exhibits multiple quantum noise suppression channels, and the noise suppression degree reaches 9.0dB when cavity detuning $\Delta_s = 0$ MHz. To realize the phase-sensitive modulation, we change the pumping light phase $\theta$ from 0 to $\pi$. In Fig. 2(e), it is clearly to see that the quadrature amplitude value changes from positive (peak) to negative (dip), and the bandwidths of multi-channel noise dips are decreased evidently compared with Fig. 2(c). In contrast, the quadrature phase value changes from negative (dip) to positive (peak) in Fig. 2(f), and the bandwidths of multi-channel noise peaks are broadened significantly compared with Fig. 2(d). This broadband and multi-channel hybrid cavity scheme has high efficiency and applicability in quantum information processing [29], weak light signal measurement [7], etc.

### 3.2 Squeezed vacuum field

Then consider the case that signal light is squeezed field rather than vacuum field in a single cavity (without cavity mirror $M_{mid}$). Turn off the pumping light and let the squeezed index of signal field be 0.5 [15]. Other parameters are same as Fig. 2(a). As shown in Fig. 3(a), the squeezed state is amplified compared with the original noise curve, an evident squeezed noise dip occurs at cavity resonance frequency $\Delta_s = 0$ MHz. This EIT-like phenomenon is generated by the strong coupling between the squeezed filed and the degenerate OPA mode. In contrast, the quadrature phase shows an amplified state at $\Delta_s = 0$ MHz in Fig. 3(b). Two symmetrical squeeze dips appear at two sideband frequencies, and the noise suppression degrees are 3.9dB.

Fig. 3(c) and Fig. 3(d) show the noise properties in a hybrid resonant cavity (with cavity mirror $M_{mid}$). The pumping light is turned on and $\theta = 0$. In Fig. 3(c), the bandwidths of resonant noise amplified state ($\Delta_s = 0$ MHz) and two sideband amplified states ($\Delta_s = \pm 3.1$ MHz) are further extended compared with Fig. 2(c). Moreover, there are three EIT-like splitting dips at three noise amplified peaks. The EIT-like phenomenon is caused by the strong coupling between independent signal modes in two cavities $C_1$ and $C_2$ and the coupling between the signal mode with the nonlinear crystal in cavity $C_2$, which lead to sideband interference effect. The interference strength is determined by the nonlinear susceptibility $g$ and the cavity-to-cavity coupling coefficient $t_2$. Three noise amplified states are enhanced as the pumping field power increases. However, the relative depth of three

EIT-like dips decrease. In Fig. 3(d), the bandwidths of three noise squeezed states are also further extended compared with Fig. 2(d). The corresponding EIT-like peaks attenuate quickly as the pumping field power increases.

Fig. 3(e) and Fig. 3(f) show the noise curve in a hybrid resonant cavity with $\theta = \pi$. In Fig. 3(e), the noise suppression depth of quadrature amplitude is enhanced greatly compared to the vacuum field case in Fig. 2(e), and the noise suppression degree reaches 11.9dB at $\Delta_s = 0$ MHz when the pumping power is $0.5\beta_{th}$. Fig. 3(f) shows that the phase modulation makes quantum noise amplification of quadrature phase with squeezed state signal field. The bandwidths of amplification peaks are broadened evidently compared to the bandwidths of squeezed dips in Fig. 3(d), and the noise amplification degree reaches 13.7dB. From the above discussion, the noise intensity can be switched from the minimum (suppression) to the maximum (amplification) or vice versa by phase-sensitive modulation. In addition, phase-sensitive modulation also broadens the output bandwidth of multi-channel signal field greatly.

In a hybrid optical cavity system, the effect of quantum noise suppression is enhanced greatly as the pumping field power increases, but the EIT-like phenomenon is attenuated simultaneously due to the absorptive saturation between the nonlinear crystal material and cavity fields. Thus, a threshold $0.65\beta_{th}$ can be obtained at which the EIT-like effect disappears. Fig. 4 plots the noise curve with and without the pumping field. In Fig. 4(a-d), the multi-channel noise intensities are switched from the minimum (suppression) to the maximum (amplification) or vice versa due to the appearance of the pumping field, i.e., the pumping field acts as a switching light. In Figs. 4(e) and 4(f), a high noise suppression degree 13.9dB of quadrature amplitude and a noise amplification degree 18.9dB of quadrature phase are obtained. As far as we know, this is the first time that multi-channel quantum noise suppression is well realizes at the quadrature amplitude of squeezed field. It can be used to various fields like precise measurement physics [10], quantum information processing [11], optical communications [12], etc.

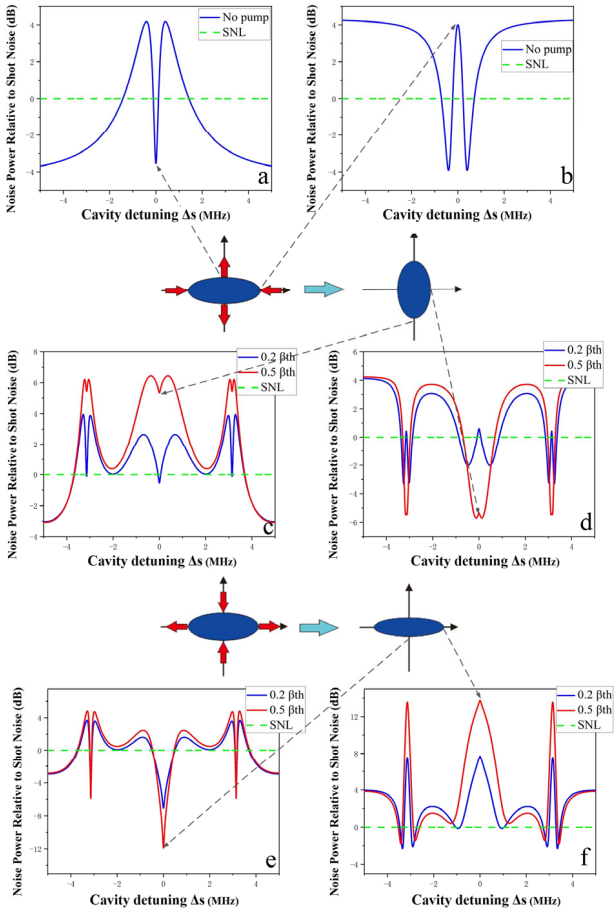

FIG. 3. The quantum correlation noise fluctuation curve of squeezed field. Other parameter values are the same as in Fig. 2.

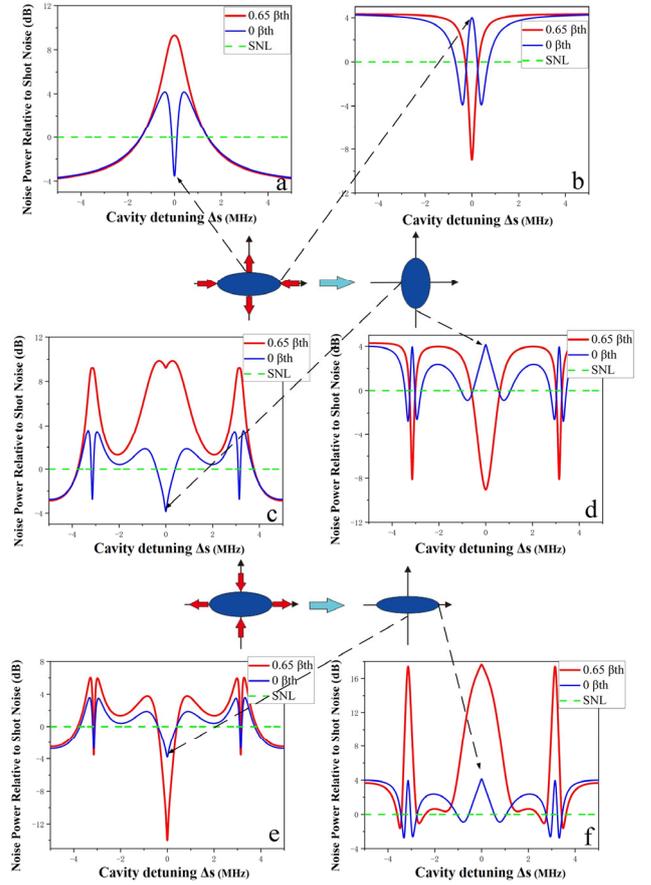

FIG. 4. The quantum noise curve with and without the pumping field in a hybrid cavity system. Green curve: the shot-noise limit (SNL); Blue curve: without the pumping field; Red curve: the pumping field power is $0.65\ \beta_{th}$. Other parameter values are same as Fig. 3.

## 4. Conclusions

In summary, the multi-channel quantum noise suppression and phase-sensitive modulation of vacuum field and CV

### 3.3 The saturation point of pumping light in squeezed vacuum field

squeezed fields in a hybrid cavity system are analyzed systematically. The simultaneous multi-channel and high noise suppression scheme significantly improves the operating efficiency and applied range of CV squeezed field. In addition, the multi-channel bandwidths and the noise intensities can be well manipulated by phase-sensitive modulation. A noise suppression degree 13.9dB of quadrature amplitude is obtained. The EIT-like phenomenon is found in noise curve due to the interference of cavity fields in the nonlinear crystal material. The effective phase-sensitive modulation of squeezed field fills the gap of quantum noise suppression in quadrature amplitude, thus our result is of great significance in various applications, especially in ultra-high-speed quantum computing [27, 28], quantum information processing [11, 12], etc.

This work was supported by National Natural Science Foundation of China (Grant Nos. 11704053, 52175531); the Science and Technology Research Program of Chongqing Municipal Education Commission (Grant No. KJQN201800629); the Innovation Leader Talent Project of Chongqing Science and Technology (Grant No. CSTC-CXLJRC201711); the Postdoctoral Applied Research Program of Qingdao (Grant No. 62350079311135); the Postdoctoral Applied Innovation Program of Shandong (Grant No. 62350070311227). the National Key Research and Development Program of China (Grant No. 2021YFC2203601).

## AUTHOR DECLARATIONS
Conflict of Interest
The authors have no conflicts to disclose.

## DATA AVAILABILITY
The data that support the findings of this study are available from the corresponding author upon reasonable request.